\documentclass[12pt]{iopart}

\pdfoutput=1

\usepackage{latexsym,amssymb}
\usepackage[english]{babel}
\usepackage{graphicx}
\usepackage[utf8]{inputenc}
\usepackage{textcomp}
\usepackage[babel=true,final=true,protrusion=true, expansion=true]{microtype}
\usepackage[colorlinks,urlcolor=blue,citecolor=blue,unicode]{hyperref}
\hypersetup{
	pdftitle={Interference effects in hybrid cavity optomechanics},
	pdfauthor={O. \v{C}ernot\'ik, C. Genes, A. Dantan}
}
\usepackage{verbatim}

\newcommand{\Hint}{\ensuremath{H_\mathrm{int}}}
\newcommand{\Hdr}{\ensuremath{H_\mathrm{dr}}}
\newcommand{\mrm}[1]{\ensuremath{\mathrm{#1}}}
\newcommand{\avg}[1]{\ensuremath{\langle #1\rangle}}
\newcommand{\nbar}{\ensuremath{\bar{n}}}
\newcommand{\inpt}{\ensuremath{\mathrm{in}}}
\newcommand{\out}{\ensuremath{\mathrm{out}}}
\newcommand{\Hc}{\ensuremath{\mathrm{H.c.}}}
\newcommand{\vect}[1]{\ensuremath{\bi{#1}}}
\newcommand{\eqref}[1]{\eref{#1}}
\newcommand{\eff}{\mrm{eff}}

\begin{document}

\title{Interference effects in hybrid cavity optomechanics}

\author{Ond\v{r}ej \v{C}ernot{\'i}k$^1$, Claudiu Genes$^1$ and Aur{\'e}lien Dantan$^2$}
\address{$^1$ Max Planck Institute for the Science of Light, Staudtstra\ss{}e 2, 91058 Erlangen, Germany}
\address{$^2$ Department of Physics and Astronomy, University of Aarhus, DK-8000 Aarhus C, Denmark}
\ead{ondrej.cernotik@mpl.mpg.de}

\begin{abstract}
    Radiation pressure forces in cavity optomechanics allow for efficient cooling of vibrational modes of macroscopic mechanical resonators, the manipulation of their quantum states, as well as generation of optomechanical entanglement.
    The standard mechanism relies on the cavity photons directly modifying the state of the mechanical resonator.
    Hybrid cavity optomechanics provides an alternative approach by coupling mechanical objects to quantum emitters, either directly or indirectly via the common interaction with a cavity field mode.
    While many approaches exist, they typically share a simple effective description in terms of a single force acting on the mechanical resonator.
    More generally, one can study the interplay between various forces acting on the mechanical resonator in such hybrid mechanical devices.
    This interplay can lead to interference effects that may, for instance, improve cooling of the mechanical motion or lead to generation of entanglement between various parts of the hybrid device.
    Here, we provide such an example of a hybrid optomechanical system where an ensemble of quantum emitters is embedded into the mechanical resonator formed by a vibrating membrane.
    The interference between the radiation pressure force and the mechanically modulated Tavis--Cummings interaction leads to enhanced cooling dynamics in regimes in which neither force is efficient by itself.
    Our results pave the way towards engineering novel optomechanical interactions in hybrid optomechanical systems.
\end{abstract}

\date{\today}

\maketitle


\section{Introduction}

Cavity optomechanics~\cite{Aspelmeyer2014} has reached a remarkable success in coupling high-quality mechanical resonators and light via radiation pressure.
This interaction can be used for measurements of small mechanical displacements and external forces~\cite{Regal2008,Hertzberg2009,Forstner2012,Schreppler2014},
for quantum state transfer between the cavity field and the mechanical oscillator, and for ground state mechanical cooling~\cite{Chan2011,Teufel2011}.
Other achievements are frequency conversion between cavity modes~\cite{Dong2012,Hill2012,Andrews2014,Cernotik2017,Midolo2018},
generation of two-mode squeezing useful for amplification of the mechanical motion or the cavity field~\cite{Ockeloen-Korppi2016,Toth2017},
and the creation of photon--phonon or phonon--phonon entanglement~\cite{Palomaki2013b,Riedinger2016,Ockeloen-Korppi2018,Riedinger2018}.
Many of these applications rely on the simultaneous fulfilment of two requirements:
i) operating in the resolved sideband regime in which the cavity linewidth is smaller than the mechanical frequency and
ii) having a sufficiently strong coupling between photons and phonons.
In systems based on optical Fabry--P\'erot resonators (such as membrane-in-the-middle optomechanical devices~\cite{Thompson2008,Jayich2008}), these two conditions are not independent;
using a short optical cavity (leading to a small mode volume and large coupling strengths) results in a large cavity decay rate such that the resolved sideband regime cannot be reached.
The sideband resolution is improved by using a long cavity in which, however, the coupling is reduced owing to the large mode volume.
It is therefore desirable to investigate alternative approaches that can either relax the conditions on sideband resolution or improve the coupling strength without increasing the decay rate.

In recent years, hybrid optomechanical systems emerged as an interesting platform for novel optomechanical experiments~\cite{Treutlein2014,Kurizki2015}.
In these systems, cavity fields interact with mechanical oscillators and few-level systems, such as single atoms or their ensembles~\cite{Tian2004,Hammerer2009b,Hammerer2010,Camerer2011,Restrepo2014,Restrepo2016}, Bose--Einstein condensates~\cite{Treutlein2007,Hunger2010}, colour centres~\cite{Rugar2004,Arcizet2011,Pigeau2015,Golter2016}, or superconducting circuits~\cite{Etaki2008,LaHaye2009,Stannigel2010,Pirkkalainen2013,Abdi2015,Cernotik2016}.
For instance, interaction with an atomic ensemble can lead to backaction evading measurements of mechanical motion~\cite{Bariani2015,Moller2017},
 generation of entanglement between the ensemble and mechanical oscillator~\cite{Genes2008,Hammerer2009a,Huang2018},
or cooling of the mechanical motion in the unresolved sideband regime~\cite{Genes2009,Genes2011,Jockel2015}.

The interplay of various types of interactions in hybrid quantum systems can lead to interference effects and novel optomechanical phenomena.
Several works have pointed out the role of interference in standard and hybrid optomechanics~\cite{Elste2009,Genes2009,Xuereb2011,Genes2011,Sawadsky2015,Yanay2016} and shown it to be decisive in obtaining, for example, novel, efficient forms for optomechanical cooling.
A particularly interesting situation arises when a vibrating membrane is doped by an ensemble of two-level emitters as shown schematically in figure~\ref{fig:scheme}(a).
Such a setup has been investigated for the first time in Ref.~~\cite{Dantan2014} where a poorly reflecting membrane oscillator was considered.
Radiation pressure forces thus played a negligible role but, owing to the presence of the dopant, the oscillator experienced an effective optomechanical interaction with the cavity mode.
Such a coupling allows for efficient optomechanical cooling in the unresolved cavity limit, enabled by dressing of the cavity field by the narrow-linewidth emitters.
A legitimate question, potentially relevant for a wide range of hybrid optomechanical systems, concerns the interplay between this position-modulated Tavis--Cummings interaction and radiation pressure when the mechanical resonator is partially reflecting and radiation pressure can no longer be neglected.

In this work we theoretically investigate the optomechanical effects arising from these two types of interaction.
The presence of the dopant results in a Fano resonance in the cavity noise spectrum which can be used to suppress the Stokes scattering (responsible for heating of the mechanical motion) and enhance the anti-Stokes scattering (cooling), leading to improved cooling performance.
Radiation pressure can further boost this effect such that the resulting optomechanical forces lead to stronger optomechanical cooling of the mechanics, as compared to the situations in which either the dopant-induced optomechanical force or radiation pressure acts independently.
In particular, we demonstrate that efficient cooling is achievable in situations in which neither dopant-induced nor radiation pressure cooling perform well.
We focus on the case of a bad optomechanical cavity---a short cavity containing a movable membrane~\cite{Flowers2012,Shkarin2014}---in which a large optomechanical coupling can be achieved, but the bare cavity linewidth is too large to resolve the mechanical sidebands.
To make the discussion simple we focus on the case of a partially reflecting membrane doped with two-level systems that interact with the cavity field via a Tavis--Cummings interaction.
Our results could, however, be amenable to other hybrid mechanical resonators doped with single or multiple two-level emitters (such as diamond cantilevers~\cite{Arcizet2011,Kolkowitz2012,Barfuss2015}, nanowires~\cite{Yeo2013}, optically or electrically trapped nanospheres~\cite{Kuhlicke2014,Delord2017}, or photonic crystals~\cite{Cotrufo2017}) and illustrate how interference effects can be exploited for engineering of efficient optomechanical interactions in hybrid mechanical systems.

\section{Model}\label{sec:model}

\begin{figure}
    \centering
    \includegraphics[width=0.6\linewidth]{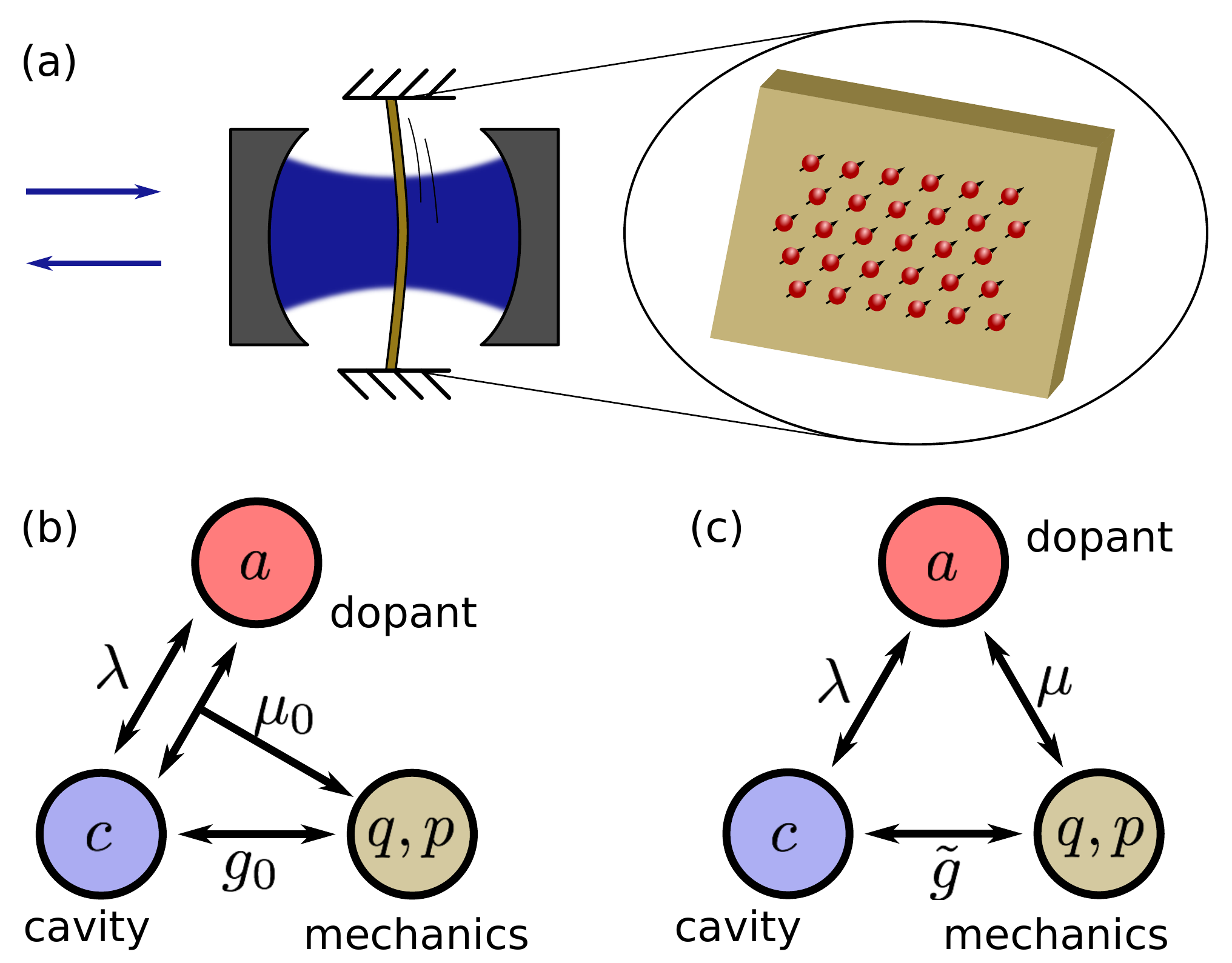}
    \caption{\label{fig:scheme}
    (a)~Schematic of the setup.
    We consider a cavity optomechanical system in the membrane-in-the-middle configuration;
    the membrane is doped with an ensemble of two-level emitters that collectively behave as a single bosonic mode.
    (b)~Depiction of the interactions of the three modes in the fundamental nonlinear configuration given by Hamiltonian~\eqref{eq:H_nonlin}.
    (c)~Interactions in the linearized regime as described by the Hamiltonian~\eqref{eq:H_lin}.}
\end{figure}

We consider the system depicted in figure~\ref{fig:scheme} where a single cavity mode $c$ interacts with a single vibrational mode of a flexible membrane with an embedded ensemble of two-level quantum emitters.
Following Ref.~\cite{Dantan2014}, we consider the limit of weak excitation of the ensemble, such that its collective spin can be described by the bosonic annihilation operator $a$ (with the commutator $[a,a^\dagger] = 1$).
The system then follows the Hamiltonian
\begin{equation}\label{eq:H_nonlin}
    H = H_0 + \Hint + \Hdr.
\end{equation}
The bare Hamiltonian $H_0/\hbar = \omega_{c} c^\dagger c + \omega_{a} a^\dagger a + \omega_\mrm{m}(q^2+p^2)/2$ describes the free evolution of the cavity field at frequency $\omega_{c}$, the dopant spin at frequency $\omega_{a}$, and the mechanical resonator with displacement $q$ and momentum $p$ (obeying the commutation relation $[q,p] = \rmi$) at frequency $\omega_\mrm{m}$.
The last term, $\Hdr/\hbar = -\rmi\eta\, c\exp(\rmi\omega_\mrm{L}t+\rmi\phi)+\Hc$, describes driving of the cavity mode with laser light of frequency $\omega_\mrm{L}$, amplitude $\eta$, and phase $\phi$.

The interaction Hamiltonian describes the interaction of the cavity field with the mechanical oscillator via radiation pressure and with the dopant via a mechanically modulated Tavis--Cummings coupling~\cite{Dantan2014}, $\Hint/\hbar = g_0c^\dagger cq + (\lambda + \mu_0q)(a^\dagger c + c^\dagger a)$;
cf. figure~\ref{fig:scheme}(b).
Here, the displacement dependence of the Tavis--Cummings interaction arises from the motion of the membrane which shifts the position of the dopant in the standing wave of the cavity mode; for a membrane placed in the middle between a node and an antinode of the field and dopant in the Lamb--Dicke regime, expansion to the first order in mechanical displacement is sufficient to characterize all dynamical effects~\cite{Dantan2014}.
The coupling $\lambda=\sqrt{N} d\sqrt{\omega_c/(2\epsilon_0 \hbar l S_{\eff})}$ stems from the collective interaction of $N$ emitters (with individual dipole moment $d$) with the zero-point field amplitude of the cavity (inversely proportional to the square root of the quantization volume $l S_{\eff}$).
We assume a Fabry--Perot type of cavity of length $l$, finesse $\cal{F}$, mode area $S_{\eff}$ and resulting mode linewidth $\kappa=\pi c_0/l \cal{F}$ ($c_0$ is the speed of light).
The dipole moment of a dopant emiter is of course directly related to the spontaneous emission rate $\gamma=\omega_a^3d^2/(3\pi \epsilon_0\hbar c_0^3)$ such that the ratio $\lambda^2/\gamma$ does not depend on the choice of emitter.
We can then estimate that the dopant--cavity cooperativity $C=N\lambda^2/(\kappa\gamma)=3N{\cal{F}}(\lambda_c/2\pi)^2/S_{\eff}$ depends mainly on the cavity design and number of dopant atoms; here, $\lambda_c = 2\pi c_0/\omega_c$ is the cavity wavelength.
In the remainder of the article, we put $\hbar = 1$ for simplicity.

\subsection{Linearized dynamics}

We linearize the Hamiltonian~\eqref{eq:H_nonlin} using the standard approach outlined in detail in \ref{app:linearization}.
We start by formulating and solving the classical equations of motion of the system in the steady state.
Provided a single steady state solution exists with solutions $\bar{c}$, $\bar{a}$, $\bar{q}$ (i.e., the system is statically stable), we formulate linearized equations of motion for the quantum fluctuations around this steady state, $c = \bar{c}+\delta c$, $a = \bar{a}+\delta a$, $q = \bar{q}+\delta q$.
Depending on the strength of the interactions, these linearized equations might become dynamically unstable; we defer discussion of dynamical stability to section~\ref{ssec:numerics}.
Assuming the stability criteria are met, the linearization procedure yields the Hamiltonian $H = H_0 + \Hint$, where
\begin{equation}
\eqalign{\label{eq:H_lin}
    H_0 = \Delta_{c} c^\dagger c + \Delta_{a} a^\dagger a + \frac{\omega_\mrm{m}}{2}(q^2+p^2),\\
    \Hint = (\tilde{g}^\ast c + \tilde{g} c^\dagger)q + \lambda(a^\dagger c + c^\dagger a) + \mu(a+a^\dagger)q.}
\end{equation}
Here, $\Delta_i = \omega_i - \omega_\mrm{L}$ is the detuning of the respective mode ($i = {a},{c}$) from the laser drive frequency and
\begin{equation}
    \tilde{g} = g - \frac{\rmi\lambda\mu}{\gamma + \rmi\Delta_{a}};
\end{equation}
we also defined the linearized coupling rates $g = g_0\bar{c}$, $\mu = \mu_0\bar{c}$.
Notice that we have dropped the $\delta$ and simply denote the fluctuations by $c$, $a$, $q$ for simplicity.
A simplified diagram of the interactions in the linearized regime is depicted in figure \ref{fig:scheme}(c).

In this linearized regime, the dynamics of the mechanical oscillator are given by the Langevin equations
\begin{equation}\label{eq:LangevinMech}
    \dot{q} = \omega_\mrm{m}p,\qquad \dot{p} = -\omega_\mrm{m}q - \gamma_\mrm{m}p + \xi - F,
\end{equation}
where $\gamma_\mrm{m}$ is the intrinsic mechanical linewidth and $\xi$ the associated bath operator;
it has zero mean and correlation function $\avg{\xi(t)\xi(t')} = \gamma_\mrm{m}(2\nbar + 1)\delta(t-t')$ with the average thermal phonon number $\nbar$.
In addition to the thermal bath $\xi$, the mechanical resonator is also coupled to an effective bath represented by a zero-average noise term with contributions from the atomic and cavity degrees of freedom
\begin{equation}\label{eq:LangevinForce}
    F = \tilde{g}^\ast c + \tilde{g} c^\dagger +\mu(a+a^\dagger).
\end{equation}
To describe the properties of this extra Langevin noise term, we list the equations of motion for the cavity field and the dopant,
\numparts
\begin{eqnarray}
    \dot{c} = -(\kappa+\rmi\Delta_{c})c - \rmi\tilde{g}q - \rmi\lambda a + \sqrt{2\kappa}c_\inpt,\\
    \dot{a} = -(\gamma+\rmi\Delta_{a})a - \rmi\mu q - \rmi\lambda c + \sqrt{2\gamma}a_\inpt.
\end{eqnarray}
\endnumparts
The cavity field decays at a rate $\kappa$, is driven by the noise operator $c_\inpt$ with zero mean and correlation function $\avg{c_\inpt(t)c_\inpt^\dagger(t')} = \delta(t-t')$, and its output follows the relation $c_\out = \sqrt{2\kappa}c - c_\inpt$.
Analogous relations hold also for the dopant for which the decay rate $\kappa$ is replaced by $\gamma$.

To quantify the effect of the extra Langevin noise term \eqref{eq:LangevinForce} on the dynamics of the mechanical resonator, we follow a perturbative approach~\cite{Marquardt2007,Genes2008a} in which we ignore the backaction of the mechanical resonator on the field and dopant.
To zeroth order in the mechanical displacement $q$, the cavity field and the dopant ensemble in frequency space can be expressed as
\numparts
\begin{eqnarray}
    c(\omega) = \tilde{\chi}_{c}(\omega)[\sqrt{2\kappa}c_\inpt - \rmi\lambda\chi_{a}(\omega)\sqrt{2\gamma}a_\inpt],\\
    a(\omega) = \tilde{\chi}_{a}(\omega)[\sqrt{2\gamma}a_\inpt - \rmi\lambda\chi_{c}(\omega)\sqrt{2\kappa}c_\inpt],
\end{eqnarray}
\endnumparts
where we introduced the bare and dressed susceptibilities
\begin{eqnarray}\eqalign{
    \chi_{c}^{-1}(\omega) = \kappa - \rmi(\omega-\Delta_{c}),\qquad
    \tilde{\chi}_{c}^{-1}(\omega) = \chi_{c}^{-1}(\omega) + \lambda^2\chi_{a}(\omega)\\
    \chi_{a}^{-1}(\omega) = \gamma - \rmi(\omega-\Delta_{a}),\qquad
    \tilde{\chi}_{a}^{-1}(\omega) = \chi_{a}^{-1}(\omega) + \lambda^2\chi_{c}(\omega).}
\end{eqnarray}
With these solutions, we can rewrite the Langevin force as
\begin{eqnarray}\fl
    F = [\tilde{g}^\ast\tilde{\chi}_{c}(\omega)-\rmi\lambda\mu\tilde{\chi}_{a}(\omega)\chi_{c}(\omega)]\sqrt{2\kappa}c_\inpt
    + [\mu\tilde{\chi}_{a}(\omega)-\rmi\tilde{g}^\ast\lambda\tilde{\chi}_{c}(\omega)\chi_{a}(\omega)]\sqrt{2\gamma}a_\inpt + \Hc
\end{eqnarray}
We express the spectrum of the Langevin force as $S_F(\omega) = S_\kappa(\omega) + S_\gamma(\omega)$ with
\numparts
\begin{eqnarray}
    S_\kappa(\omega) = 2\kappa|\tilde{g}^\ast\tilde{\chi}_{c}(\omega)-\rmi\lambda\mu\tilde{\chi}_{a}(\omega)\chi_{c}(\omega)|^2,\label{eq:Skappa}\\
    S_\gamma(\omega) = 2\gamma|\mu\tilde{\chi}_{a}(\omega)-\rmi\tilde{g}^\ast\lambda\tilde{\chi}_{c}(\omega)\chi_{a}(\omega)|^2.\label{eq:Sgamma}
\end{eqnarray}
\endnumparts
Using the force spectrum, we obtain the cooling rate~\cite{Marquardt2007}
\begin{equation}\label{eq:GammaCool}
    \Gamma_\mathrm{cool} = \frac{1}{2}[S_F(\omega_\mrm{m}) - S_F(-\omega_\mrm{m})]
\end{equation}

\subsection{Overview of cooling strategies}

\begin{figure}
    \centering
    \includegraphics[width=0.8\linewidth]{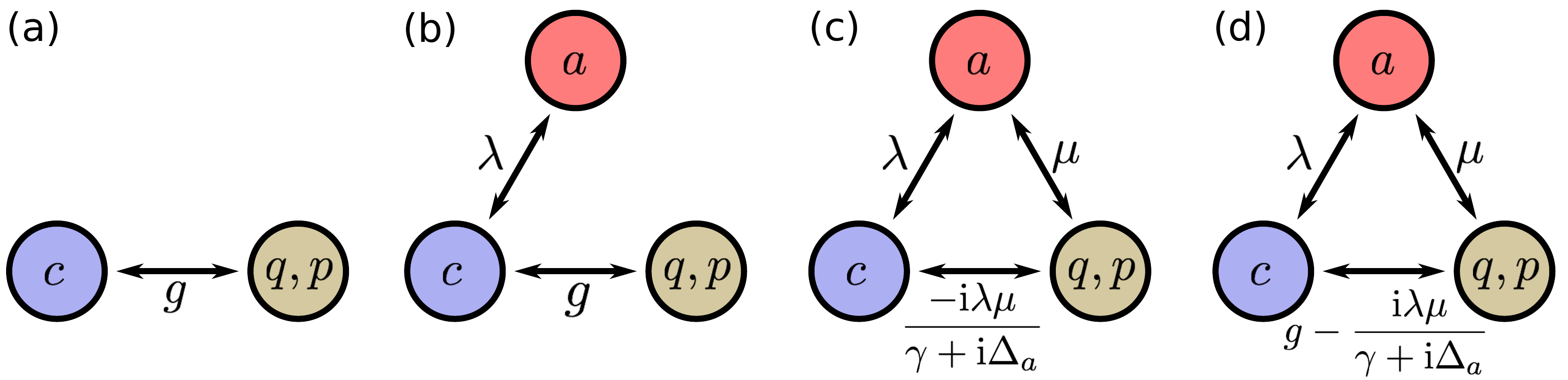}
    \caption{\label{fig:cooling}
    Schematic depiction of the interactions involved in existing cooling schemes.
    (a) Radiation pressure cooling~\cite{Marquardt2007,Wilson-Rae2007}.
    (b) Dressed cavity cooling~\cite{Genes2009}.
    (c) Dopant cooling~\cite{Dantan2014}.
    (d) Interference cooling discussed in this article.}
\end{figure}

We can use the noise spectra \eqref{eq:Skappa}, \eqref{eq:Sgamma} to recover existing approaches to optomechanical cooling.
First, the standard sideband cooling strategy~\cite{Marquardt2007,Wilson-Rae2007} corresponds to $\lambda = \mu = 0$; cf. figure~\ref{fig:cooling}(a).
In this case, we get the Lorentzian cavity spectrum $S_\kappa(\omega) = 2g^2\kappa / [\kappa^2 + (\omega - \Delta_{c})^2]$ while $S_\gamma(\omega) = 0$.
It then follows that the best cooling can be achieved with a sideband resolved system, $\kappa<\omega_\mrm{m}$, driven on the red sideband, $\Delta_{c} = \omega_\mrm{m}$;
final mechanical occupation smaller than unity requires strong optomechanical cooperativity $g^2/\kappa\gamma\nbar > 1$.
In the following, we will refer to this strategy as \emph{radiation pressure cooling}.

In the bad cavity regime, $\kappa > \omega_\mrm{m}$, radiation pressure cooling cannot reach the quantum ground state of the mechanical resonator.
To suppress the unwanted Stokes scattering in this situation, one can use an atomic ensemble placed within the same optical cavity.
If the atoms are in the resolved sideband regime, $\gamma<\omega_\mrm{m}$, they will burn a hole in the cavity spectrum;
by choosing a suitable set of detunings $\Delta_{c}$, $\Delta_{a}$, this spectral hole can overlap with the location of the upper mechanical sideband.
This modification results in a reduced density of states around the sideband, leading to reduced Stokes scattering such that the mechanical ground state can be reached~\cite{Genes2009}.
This strategy, which we will call \emph{dressed cavity cooling}, corresponds to the limit $\mu = 0$ and is shown in figure~\ref{fig:cooling}(b).

Finally, the regime with $g = 0$ [see also figure~\ref{fig:cooling}(c)] has been studied in Ref.~\cite{Dantan2014};
this situation describes a doped membrane with negligible radiation pressure coupling.
Here, the dopant provides both the sideband resolution (when in the regime $\gamma<\omega_\mrm{m}$) and coupling to the mechanical resonator (via the coupling constant $\mu$).
The cavity field (which does not couple to the mechanical motion directly) serves only to enhance the intrinsically weak interaction between the dopant and the mechanical resonator.
We name this strategy \emph{dopant cooling}.

In contrast, we investigate a cooling strategy where all three interactions---radiation pressure coupling at a rate $g$, Tavis--Cummings interaction at a rate $\lambda$, and dopant--mechanical coupling at a rate $\mu$---are present in the system at the same time (see also figure~\ref{fig:cooling}(d)).
This situation might appear identical to the dopant cooling shown in figure~\ref{fig:cooling}(c) but these two schemes differ in the effective optomechanical coupling.
While the effective optomechanical coupling in the dopant cooling scheme is $-\rmi\lambda\mu/(\gamma+\rmi\Delta_{a})$, it is equal to $\tilde{g} = g -\rmi\lambda\mu/(\gamma+\rmi\Delta_{a})$ in our model.
This latter form of the coupling leads to detuning-dependent interference between radiation pressure and dopant coupling which can further lower the final occupation.
Owing to this effect, we denote this strategy \emph{interference cooling}.

\subsection{Fano resonance}

The noise spectra \eqref{eq:Skappa}, \eqref{eq:Sgamma} reveal that interference can play an important role in cooling.
For example, the cavity input noise can influence the mechanical motion either directly from the cavity field (dressed by the presence of the dopant),
or it can be transferred to the dopant and affect the motion from there.
These processes are captured by the first and second term in \eqref{eq:Skappa}, respectively;
since they both stem from the same reservoir, they have to be added coherently.
Different interference conditions exist for the Stokes and anti-Stokes scattering, such that strong asymmetry in the two processes is possible even in the bad cavity regime.

In the following, we will consider cooling in the bad cavity regime, $\kappa \gg \omega_\mrm{m}$, and assume that the dopant is relatively strongly coupled to the cavity field such that the Tavis--Cummings interaction is in the regime of strong cooperativity, $C=\lambda^2/\kappa\gamma > 1$.
From the scaling $C=3N{\cal{F}}(\lambda_c/2\pi)^2/S_{\eff}$ derived in section~\ref{sec:model}, one can estimate that in order to reach this regime, a cavity of mode area around $50\mu$m$\times50\mu$m would require the product $N{\cal{F}}$ to achieve values of the order $10^6$.
As cavities of finesse around $10^4-10^5$ are routinely employed in optomechanical setups, one then requires around $100$ dopant emitters to be placed in the mode area of the cavity field, thus at relatively low densities (such that the emitters can be considered independent).

The cavity mode and the dopant form polaritons with energies
\begin{equation}
    \omega_\pm = \frac{1}{2}\left[\Delta_{a}+\Delta_{c} \pm \sqrt{(\Delta_{a}-\Delta_{c})^2+4\lambda^2}\right].
\end{equation}
We can expect the cooling to be optimal when one of the polariton modes is driven on the lower mechanical sideband, $\omega_+ = \omega_\mrm{m}$ (or $\omega_- = \omega_\mrm{m}$),
which is achieved for the cavity detuning
\begin{equation}\label{eq:DcOpt}
    \Delta_{c} = \omega_\mrm{m} + \frac{\lambda^2}{\Delta_{a}-\omega_\mrm{m}}.
\end{equation}

Plugging the optimal detuning~\eqref{eq:DcOpt} into the noise spectra \eqref{eq:Skappa}, \eqref{eq:Sgamma} and assuming $\kappa\gg\omega_\mrm{m}$, we can approximate the noise spectra to leading order in $\omega_\mrm{m}/\kappa$ as
\begin{equation}
    S_\kappa(\omega) \simeq \frac{A(\omega)}{\Gamma^2 + (\omega-\Delta)^2},\label{eq:SkappaFano}\qquad
    S_\gamma(\omega) \simeq \frac{B}{\Gamma^2 + (\omega-\Delta)^2}.\label{eq:SgammaLorentz}
\end{equation}
The spectra describe the hybridization of the cavity mode with the dopant; the emergent polaritonic state is characterized by linewidth $\Gamma$ and the state energy quantified by $\Delta$:
\numparts
\begin{eqnarray}\label{eq:Gamma}
    \Gamma = \frac{\lambda^4\gamma + \kappa(\lambda^2+\gamma\kappa)(\Delta_{a}-\omega_\mrm{m})^2}
        {\lambda^4+\kappa^2(\Delta_{a}-\omega_\mrm{m})^2},\\
    \Delta = \frac{\lambda^4\omega_\mrm{m} + \kappa^2\Delta_{a}(\Delta_{a}-\omega_\mrm{m})^2}
        {\lambda^4+\kappa^2(\Delta_{a}-\omega_\mrm{m})^2}.\label{eq:Delta}
\end{eqnarray}
\endnumparts
Note that, for weak coupling, one reproduces the expected result that the polariton exhibits the bare linewidth of the dopant $\Gamma\simeq \gamma$ and is positioned at $\Delta \simeq \Delta_a$. For increasing coupling strength the polariton linewidth and energy acquire contributions from both the dopant and the cavity mode.
Furthermore, the spectra \eqref{eq:SkappaFano} are characterized by the amplitudes
\numparts
\begin{eqnarray}\fl
    A(\omega) = \frac{2\kappa(\Delta_{a}-\omega_\mrm{m})^2}{\gamma^2+\Delta_{a}^2}
        \frac{\{\lambda\mu(2\Delta_{a}-\omega)-g[\gamma^2-\Delta_{a}(\omega-\Delta_{a})]\}^2 + g^2\gamma^2\omega^2}
            {\lambda^4 + \kappa^2(\Delta_{a}-\omega_\mrm{m})^2},\label{eq:amplitudeA}\\
    \fl\eqalign{
    B = \frac{2\gamma}{\gamma^2+\Delta_{a}^2}&
        \Bigg(\frac{[\lambda^2\mu\gamma - (g\lambda\gamma+\mu\kappa\Delta_{a})(\Delta_{a}-\omega_\mrm{m})]^2}
            {\lambda^4 + \kappa^2(\Delta_{a}-\omega_\mrm{m})^2}\\
        &\quad+\frac{[\lambda^2\mu(2\Delta_{a}-\omega_\mrm{m}) - (g\lambda\Delta_{a}-\mu\gamma\kappa)(\Delta_{a}-\omega_\mrm{m})]^2}
            {\lambda^4 + \kappa^2(\Delta_{a}-\omega_\mrm{m})^2}\Bigg).}
\end{eqnarray}
\endnumparts
The amplitude $A(\omega)$ is quadratic in frequency so the cavity noise spectrum $S_\kappa(\omega)$ exhibits a Fano resonance~\cite{Fano1961};
the atomic noise spectrum $S_\gamma(\omega)$, on the other hand, is Lorentzian.
The Fano resonance can be further enhanced by the interference between the radiation pressure and the dopant interaction as we discuss below.

\section{Interference cooling}\label{sec:cooling}

\subsection{Dopant-induced cooling}\label{ssec:dopant}

First, we turn our attention to the Lorentzian noise spectrum of the dopant $S_\gamma(\omega)$.
It follows from the theory of sideband cooling~\cite{Marquardt2007,Wilson-Rae2007} that the optimum cooling performance is achieved for $\Delta = \omega_\mrm{m}$ and $\Gamma < \omega_\mrm{m}$.
These conditions can be realized using a good dopant $\gamma < \omega_\mrm{m}$ with detuning $\Delta_{a} = \omega_\mrm{m}$.
The noise spectrum then simplifies to
\begin{equation}
    S_\gamma(\omega) = \frac{2\mu^2\gamma}{\gamma^2 + (\omega-\omega_\mrm{m})^2}
\end{equation}
while the cavity noise spectrum becomes zero, $S_\kappa(\omega) = 0$.
This result is quite natural, since driving the dopant on the red mechanical sideband results [for polariton driving according to \eqref{eq:DcOpt}] in an infinite cavity detuning.
The cavity is thus strongly off-resonant so it decouples from the dynamics which thus obey the Hamiltonian
\begin{equation}
    H = \frac{\omega_\mrm{m}}{2}(2a^\dagger a + q^2 + p^2) + \mu(a+a^\dagger)q.
\end{equation}

One might expect that ground state cooling in this regime is possible provided the system exhibits strong cooperativity, $\mu^2/\gamma\gamma_\mrm{m}\nbar > 1$.
This assertion is true in principle, but such a regime would be extremely difficult to reach in an experiment.
Recall that the the coupling rate $\mu = \mu_0\bar{c}$ is obtained from the three-body interaction $\mu_0(a^\dagger c+ c^\dagger a)q$ enhanced by a strong intracavity amplitude $\bar{c}$.
The three-body coupling strength $\mu_0$ is, in turn, a perturbative correction to the Tavis--Cummings interaction in the Lamb--Dicke regime so we have $\mu_0\ll\lambda$.
Moreover, reaching a large cavity amplitude $\bar{c}$ for an effectively infinite detuning $\Delta_{c}$ would require effectively infinite driving power.

\subsection{Cooling via Fano resonance}

Analysis of the cavity noise spectrum, \eqref{eq:SkappaFano}, is more involved.
Owing to the frequency dependence of the amplitude $A(\omega)$, the cavity noise spectrum exhibits a Fano resonance, which can be used to modify the Stokes and anti-Stokes scattering rates.
While a general analysis of these spectra and optimization of the cooling is, in principle, possible, it does not bring a clear physical insight into the system dynamics.
We thus only highlight the main features of this approach and defer more detailed analysis to the next section where we study the noise spectra and final mechanical occupation numerically.

The cooling rate is given by $S_\kappa(\omega_\mrm{m})$ whereas heating by $S_\kappa(-\omega_\mrm{m})$;
to exploit the Fano resonance for suppressing heating and enhancing cooling, we would therefore like the dip of the Fano resonance to fall within the vicinity of $\omega = -\omega_\mrm{m}$ while the peak should be close to $\omega = \omega_\mrm{m}$ [see figure~\ref{fig:polaritons}(b) for an illustration].
These requirements already put certain conditions on the detuning and linewidth defined in \eqref{eq:Gamma}, \eqref{eq:Delta}.
Specifically, we need a detuning with magnitude within the mechanical sidebands, $|\Delta|\lesssim\omega_\mrm{m}$ and a linewidth that is not too large either, $\Gamma\lesssim\omega_\mrm{m}$.
At the same time, we must not forget that the dopant noise spectrum \eqref{eq:SgammaLorentz} also contributes to heating and cooling of the membrane.
Ideally, we would thus have positive detuning, $\Delta > 0$, such that $S_\gamma(\omega_\mrm{m}) > S_\gamma(-\omega_\mrm{m})$.

The suppression of Stokes scattering via Fano resonance is not unique to our system.
The same principle is also used in dressed cavity and dopant cooling~\cite{Genes2009,Dantan2014}.
In these two systems, the cavity field and atoms also form two polariton modes, resulting in cavity noise spectra analogous to \eqref{eq:SkappaFano}.
With interference cooling, however, there is an additional interference between the two types of interaction---the radiation pressure interaction at a rate $g$ and the dopant--mechanical interaction at a rate $\mu$ as exemplified by the curly bracket in \eqref{eq:amplitudeA}.
This interference can lead to a further suppression of the Stokes scattering (and enhancement of anti-Stokes scattering) and thus a lower final occupation than in any of the previous cooling schemes.

An intriguing consequence of this interference effect is the possibility of cooling with both cavity and dopant driven on resonance, $\Delta_c = \Delta_a = 0$.
In this case, the cavity and dopant noise spectra are not given by \eqref{eq:SkappaFano} (unless $\lambda = \omega_\mrm{m}$) but instead by the expressions
\numparts
\begin{eqnarray}
    S_\kappa(\omega) = \frac{2\kappa}{\gamma^2}\label{eq:SkappaRes}
    \frac{g^2\gamma^2\omega^2+(g\gamma^2+\lambda\mu\omega)^2}
        {(\lambda^2+\gamma\kappa)^2 + (\gamma^2+\kappa^2-2\lambda^2)\omega^2+\omega^4},\\
    S_\gamma(\omega) = \frac{2\gamma}{\gamma^2}
    \frac{\mu^2(\lambda^2+\gamma\kappa)^2 + \gamma^2(g\lambda+\mu\omega)^2}{(\lambda^2+\gamma\kappa)^2 + (\gamma^2+\kappa^2-2\lambda^2)\omega^2+\omega^4}.\label{eq:SgammaRes}
\end{eqnarray}
\endnumparts
(These expressions can be obtained simply by setting $\Delta_a = \Delta_c = 0$ in \eqref{eq:Skappa}, \eqref{eq:Sgamma}.)
The spectra clearly reveal the importance of interference for cooling on resonance:
only when both radiation pressure and dopant interaction are present does the numerator of each of the two spectra contain a term linear in frequency.
The spectra thus distinguish between positive and negative frequencies, resulting in a net cooling or heating effect.
Specifically, we obtain the cooling rate (recall the definition given in~\eqref{eq:GammaCool})
\begin{equation}
    \Gamma_\mathrm{cool} = \frac{4g\lambda\mu\omega_\mrm{m}(\gamma+\kappa)}{(\lambda^2+\gamma\kappa)^2+(\gamma^2+\kappa^2-2\lambda^2)\omega_\mrm{m}^2+\omega_\mrm{m}^4}.
\end{equation}
The denominator is always positive so the membrane is cooled as long as $g\lambda\mu >0$ (i.e., either none or two of the coupling rates are negative).

\subsection{Numerical simulations}\label{ssec:numerics}

To check our expectations, we perform numerical simulations of the full linearized dynamics to determine the final mechanical occupation.
To this end, we formulate a Lyapunov equation for the covariance matrix of the system.
We start by defining the quadrature operators $X_{c} = (c+c^\dagger)/\sqrt{2}$, $Y_{c} = -\rmi(c-c^\dagger)/\sqrt{2}$ (and similar for the dopant) with the commutator $[X_i,Y_j] = \rmi\delta_{ij}$.
Together with the mechanical position and momentum operators, we collect these operators into the vector $\vect{r} = (X_{c},Y_{c},X_{a},Y_{a},q,p)^T$ and define the covariance matrix with elements
\begin{equation}
    V_{ij} = \avg{r_ir_j + r_jr_i} - 2\avg{r_i}\avg{r_j}.
\end{equation}
The steady-state covariance matrix $\vect{V}$ is a solution of the Lyapunov equation
\begin{equation}\label{eq:Lyapunov}
    \vect{A}\vect{V} + \vect{V}\vect{A}^T + \vect{N} = 0
\end{equation}
with drift and diffusion matrices $\vect{A}$, $\vect{N}$;
we present these matrices and discuss the dynamical stability in \ref{app:Lyapunov}.
We obtain the mechanical occupation in the steady state from the variance of the mechanical position and momentum,
\begin{equation}
    n_\mrm{f} = \frac{1}{4}(V_{55} + V_{66} - 2).
\end{equation}
Note that since the dynamics are linear, the (initially Gaussian) state of the system remains Gaussian throughout the evolution and the covariance matrix is sufficient to fully describe the correlations in the system.

\begin{figure}
    \centering
    \includegraphics[width=\textwidth]{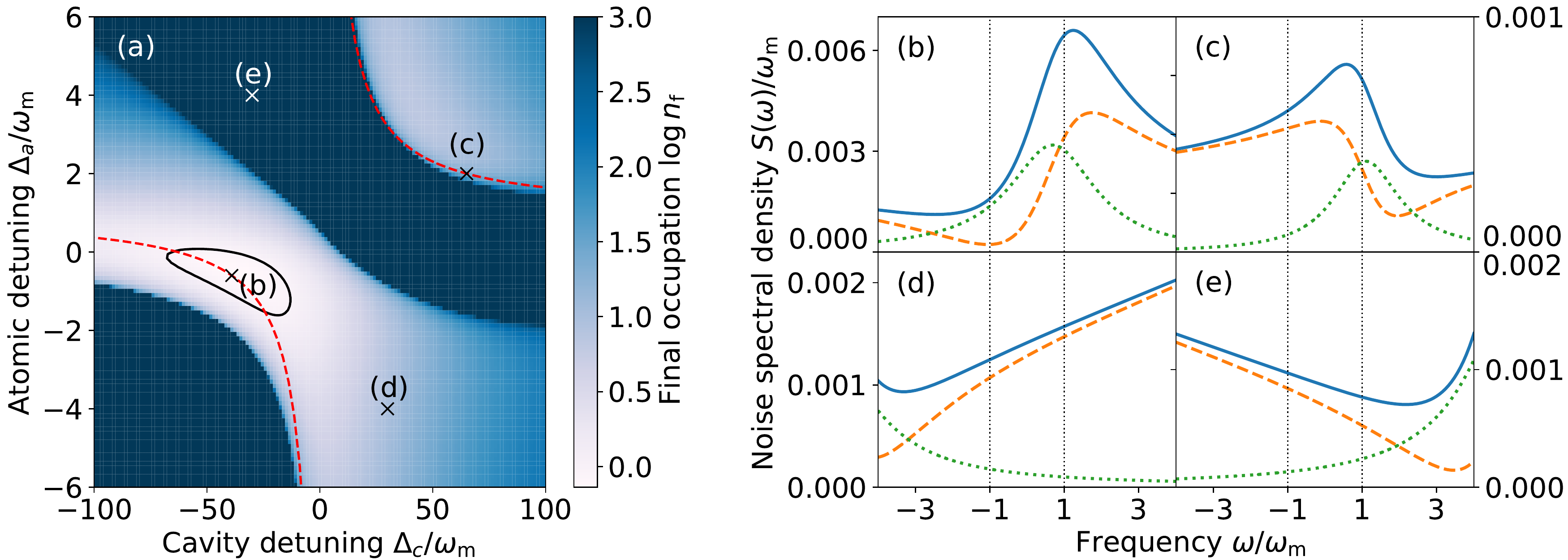}
    \caption{\label{fig:polaritons}
    (a) Final occupation of the mechanical oscillator (on logarithmic scale) as a function of the cavity and atomic detunings.
    The dashed red line shows the two polariton branches defined in \eqref{eq:DcOpt};
    the black contour line shows the region where the final occupation drops below unity, $n_\mrm{f}<1$.
    The dark blue regions are where the oscillator is heated up, $n_\mrm{f} > \nbar$, or where the system becomes unstable.
    (b--e) Noise spectral densities for detunings as indicated in panel (a).
    We show the cavity noise spectrum $S_\kappa(\omega)$ (dashed orange line), the dopant noise spectrum $S_\gamma(\omega)$ (dotted green line),
    and their sum (solid blue line).
    The system parameters are $g/\omega_\mrm{m} = 0.25$, $\lambda/\omega_\mrm{m} = 8$, $\mu/\omega_\mrm{m} = 0.01$, $\kappa/\omega_\mrm{m} = 20$, $\gamma/\omega_\mrm{m} = 0.8$, $Q_\mrm{m} = \omega_\mrm{m}/\gamma_\mrm{m} = 10^6$, and $\nbar = 10^3$.
    The vertical lines are guides to the eye for the cooling and heating rates (given by the spectra at $\omega_\mrm{m}$ and $-\omega_\mrm{m}$, respectively).}
\end{figure}

We plot the results of such a simulation in figure~\ref{fig:polaritons}(a) where we show the final occupation $n_\mrm{f}$ as a function of the cavity and dopant detunings.
Particularly, driving the upper polariton with energy $\omega_+$ on the lower mechanical sideband (shown as the dashed red line in the lower left quadrant) leads to substantive cooling and even makes it possible to reach final occupation $n_\mrm{f} < 1$.
Driving the lower polariton in the same way (upper right quadrant), on the other hand, leads only to moderate cooling or even becomes unstable (when entering the dark blue region).

We further elucidate this difference in figure~\ref{fig:polaritons}(b--e) where we plot the spectra at four different points of the 2D plot.
On the lower sideband of the upper polariton (figure~\ref{fig:polaritons}(b)), the cavity noise spectrum (dashed orange line) exhibits a clear Fano resonance which reaches a minimum around $\omega = -\omega_\mrm{m}$ and maximum close to $\omega = \omega_\mrm{m}$;
the Stokes scattering is thus suppressed while the anti-Stokes scattering is enhanced, which leads to a final occupation $n_\mrm{f}\simeq 0.74$.
On the lower sideband of the lower polariton (panel (c)), the Fano resonance is still present but not ideally oriented (the minimum is to the right of the maximum) so the final occupation is much higher ($n_\mrm{f}\simeq 19.4$).
A smaller final occupation than on the lower sideband of the lower polariton can, in fact, be achieved also far detuned from the lower sideband of the upper polariton (such as at the point (d) in figure~\ref{fig:polaritons}, where the final occupation $n_\mrm{f}\simeq 10$).
Finally, when the Stokes scattering is stronger than the anti-Stokes scattering, the system becomes unstable; cf.~figure~\ref{fig:polaritons}(e).
Together, these results reveal the importance of Fano resonance for efficient cooling:
the Fano minimum suppresses the Stokes scattering while the maximum enhances the anti-Stokes scattering.
These requirements limit the suitable dopant detuning $|\Delta_a|\lesssim \omega_\mrm{m}$ (cf. \eqref{eq:Gamma}, \eqref{eq:Delta}), leading to optimal cooling around the lower sideband of the upper polariton.

\begin{figure}[t]
    \centering
    \includegraphics[width=0.71\linewidth]{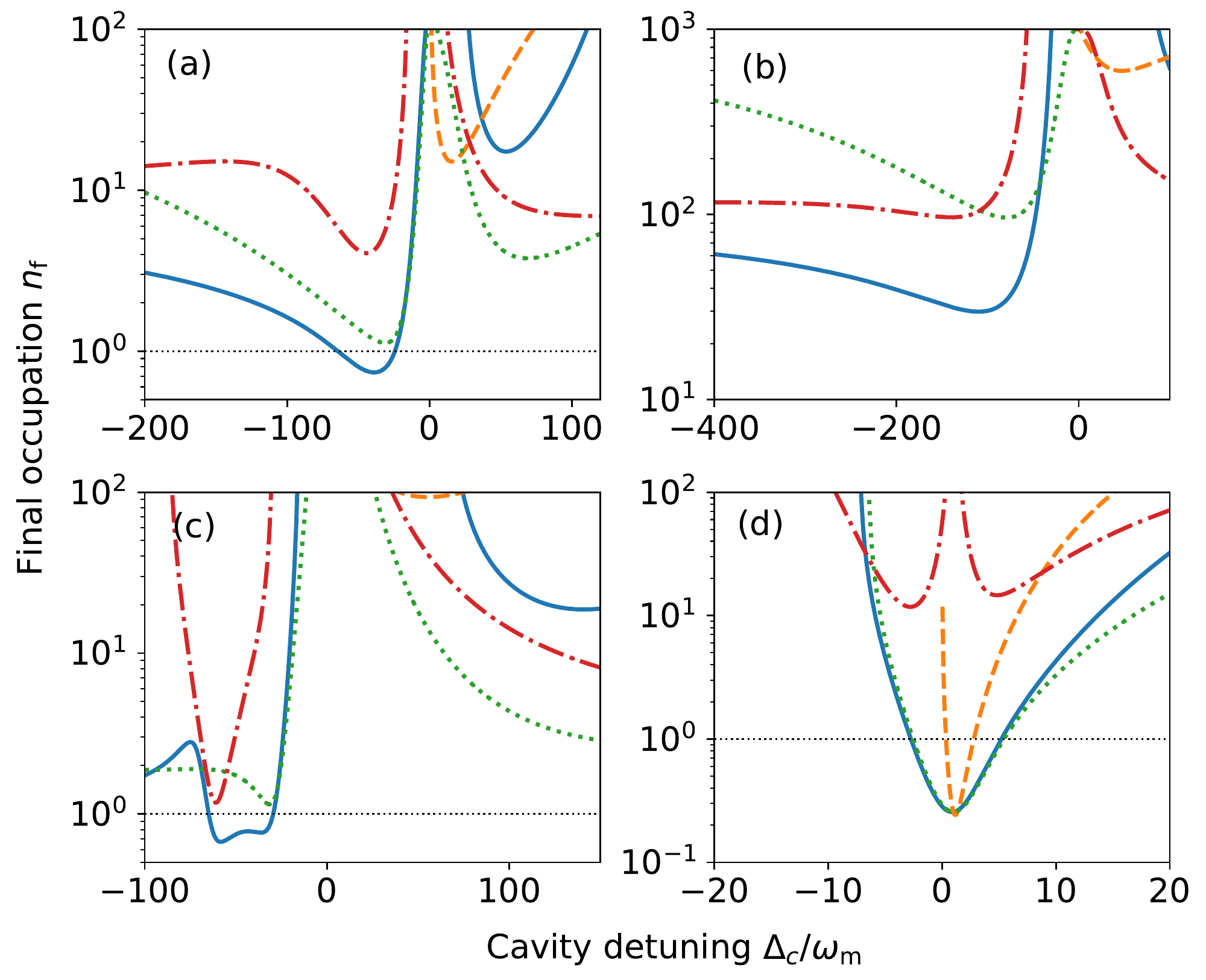}
    \caption{\label{fig:comparison}
    Comparison of cooling strategies.
    Final occupation versus cavity detuning for interference cooling (solid blue line), radiation pressure cooling (dashed orange line), dressed cavity cooling (dotted green line), and dopant cooling (dash--dotted red line) is plotted for various system parameters.
    (a) The same parameters as in figure~\ref{fig:polaritons}.
    (b) Bad cavity ($\kappa/\omega_\mrm{m} = 80$) and bad dopant ($\gamma/\omega_\mrm{m} = 2$) with coupling rates $g/\omega_\mrm{m} = 0.06$, $\lambda/\omega_\mrm{m} = 15$, $\mu/\omega_\mrm{m} = 0.006$.
    (c) Bad cavity ($\kappa/\omega_\mrm{m} = 80$) and good dopant ($\gamma/\omega_\mrm{m} = 0.1$). The coupling rates are $g/\omega_\mrm{m} = 0.3$, $\lambda/\omega_\mrm{m} = 8$, $\mu/\omega_\mrm{m} = 0.005$.
    (d) Good cavity ($\kappa/\omega_\mrm{m} = 0.8$) and bad dopant ($\gamma/\omega_\mrm{m} = 10$). Here, we use the coupling rates $g/\omega_\mrm{m} = 0.1$, $\lambda/\omega_\mrm{m} = 12$, $\mu/\omega_\mrm{m} = 0.025$.
    For interference, dressed cavity, and dopant cooling, the dopant detuning is $\Delta_{a} = \omega_\mrm{m}+\lambda^2/(\Delta_{c}-\omega_\mrm{m})$, corresponding to cooling via one of the polariton modes [i.e., along the dashed red lines in figure~\ref{fig:polaritons}(a)];
    additionally, the membrane has the mechanical quality factor $Q_\mrm{m} = 10^6$ and initial occupation $\nbar = 10^3$.
    The horizontal line indicates final occupation of unity, $n_\mrm{f} = 1$.}
\end{figure}

We study the final occupation along the lower sideband of the two polariton modes in more detail in figure~\ref{fig:comparison}(a).
Two observations are crucial here:
first, the minimum final occupation reached along the lower sideband of the upper polariton ($n_\mrm{f}\simeq 0.74$) is very close to the absolute minimum in figure~\ref{fig:polaritons} ($n_\mrm{f}\simeq 0.73$)
indicating that the lower sideband of the polariton mode is near-optimal for cooling with moderate cooperativity (we have $\lambda^2/\kappa\gamma = 4$).
Second, interference cooling (shown as the solid blue line) performs better than any other of the cooling schemes;
the best results can otherwise be achieved with dressed cavity cooling, which reaches a final occupation $n_\mrm{f}\simeq 1.1$.

We present further comparison of the four cooling schemes in figure~\ref{fig:comparison}(b--d).
There exists a broad range of system parameters---generally in the bad cavity regime---where interference cooling can outperform existing cooling strategies (panels (b,c)).
In these cases, one can reach optimum cooling for blue-detuned cavity drive, $\Delta_{c} < 0$, corresponding to rather small dopant detuning (e.g., in panel (c), the optimal dopant detuning $\Delta_{a} \simeq -\omega_\mrm{m}$).
This observation further confirms our assertion that the Fano resonance in the cavity noise spectrum is responsible for the suppression of Stokes scattering and enhancement of anti-Stokes scattering.
We also note that in the good cavity regime (panel (d)), the performance of radiation pressure, dressed cavity, and interference cooling is comparable;
admittedly, radiation pressure cooling is, from the experimental point of view, the simplest of these methods to implement.

\begin{figure}[t]
    \centering
    \includegraphics[width=0.7\linewidth]{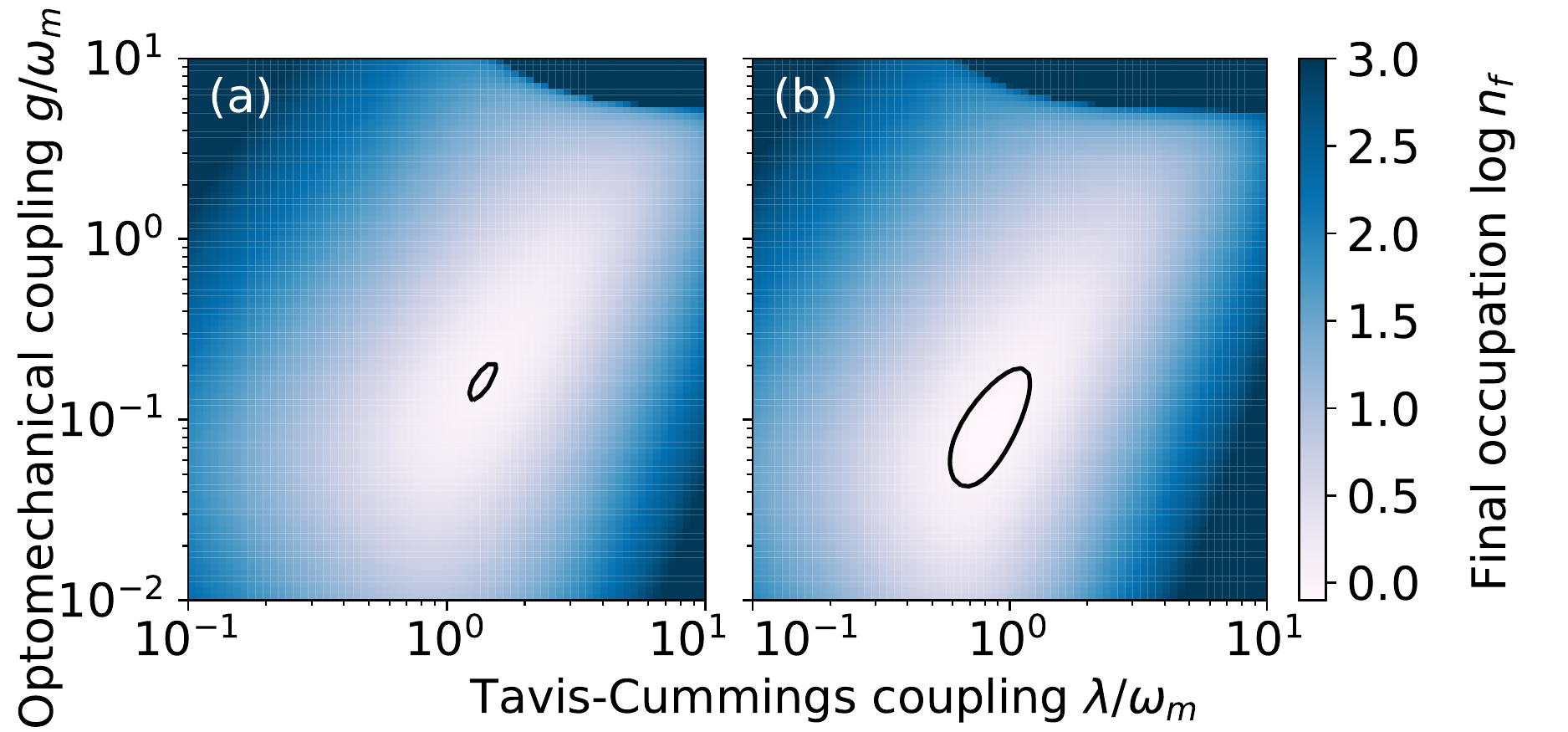}
    \caption{\label{fig:resonant}
    Final occupation (on logarithmic scale) for interference cooling with a resonant drive, $\Delta_{c} = \Delta_{a} = 0$, as a function of the Tavis--Cumings and optomechanical coupling rates in (a) the bad cavity regime ($\kappa/\omega_\mrm{m} = 2.7$, $\gamma/\omega_\mrm{m} = 0.8$) and (b) the good cavity regime ($\kappa/\omega_\mrm{m} = 0.7$, $\gamma/\omega_\mrm{m} = 0.5$).
    The black contour lines show regions where $n_\mrm{f} < 1$.
    The mechanical oscillator has the quality factor $Q_\mrm{m} = 10^6$ and initial thermal occupation $\nbar = 10^3$;
    we use the dopant coupling $\mu/\lambda = 0.05$.}
\end{figure}

Finally, we study the final occupation for interference cooling with driving on resonance, $\Delta_{c} = \Delta_{a} = 0$, in figure~\ref{fig:resonant}.
Remarkably, final occupation $n_\mrm{f} < 1$ is possible even in the bad cavity regime (panel (a)).
In the sideband resolved regime (panel (b)), the final occupation can be lower than in the bad cavity regime, but resonant interference cooling cannot outperform radiation pressure cooling;
here, the minimum final occupation is $n_\mrm{f}\simeq 0.8$ whereas radiation pressure cooling can reach $n_\mrm{f}\simeq 0.14$ with the same sideband resolution.
Nevertheless, resonant driving (as used for interference cooling) requires smaller driving power than a sideband drive (necessary for radiation pressure cooling) to achieve the same coupling strength;
interference cooling might thus have an important advantage over radiation pressure cooling even in the good cavity regime.

In both regimes, there is an optimal range of coupling rates for which ground state cooling is possible.
This effect is a consequence of the interference in the noise spectra \eqref{eq:SkappaRes}, \eqref{eq:SgammaRes}.
Take, for instance, the cavity noise spectrum $S_\kappa(\omega)$ (note, however, that a similar argument holds also for the dopant spectrum $S_\gamma(\omega)$):
here, the second term in the numerator, $\propto (g\gamma^2+\lambda\mu\omega)^2$, is responsible for cooling.
More specifically, it is the term linear in frequency, $\propto g\lambda\mu\omega$ that gives rise to cooling;
the remaining two terms, $\propto g^2, \lambda^2\mu^2$ do not affect the cooling rate but still affect the final occupation since they contribute to the backaction that the cavity field and dopant exert on the mechanical oscillator.
We therefore need to maximize the interference term relative to the latter two, which gives rise to an optimal range of coupling rates as can be seen in figure~\ref{fig:resonant}.

\section{Summary and outlook}\label{sec:conclusion}

In conclusion, we investigated cooling of a mechanical resonator doped by an ensemble of two-level quantum emitters.
The interplay between radiation pressure and mechanically modulated Tavis--Cummings interaction between the cavity field and the dopant gives rise to a Fano resonance in the cavity noise spectrum.
This resonance can lead to a suppression of Stokes and enhancement of anti-Stokes scattering, leading to ground state cooling in regimes where none of the effects alone can efficiently cool the motion.
An additional signature of the interference between these two types of interaction is the possibility of ground state cooling when the cavity and dopant are driven on resonance.
Our results are not limited to the particular architecture considered here;
similar results can be expected for any mechanical oscillator with embedded two-level quantum emitters and experiencing a direct radiation pressure force.

This work highlights the importance of interference effects in hybrid optomechanical systems for studying novel phenomena and developing new applications.
The interference can also result in a lowered instability threshold, which can have profound implications for the generation of ponderomotive squeezing of light~\cite{Fabre1994} or for observing mechanical limit cycles~\cite{Qian2012,Lorch2014}.
Further improvements and new effects may occur when the dopant ensemble is prepared in a super- or subradiant state~\cite{Plankensteiner2017} or with quadratic optomechanical coupling~\cite{Thompson2008,Jayich2008}.

Looking forward, these devices will enter a new domain once they reach the regime of near-unit reflectivity around the dopant resonance~\cite{Bettles2016,Shahmoon2017,Zeytinoglu2017a,Back2018,Scuri2018}.
Such membranes could then be used as end mirrors in Fabry--P\'erot resonators, where
their strongly frequency dependent reflectivity can reduce the cavity linewidth~\cite{Naesby2018}
and lead to the observation of non-Markovian optomechanical dynamics in the resolved sideband and strong coupling regimes.


\ack    We gratefully acknowledge financial support from the Max Planck Society and the Velux Foundations.

\appendix

\section{Linearization of the three-body dynamics and static stability}\label{app:linearization}

For completeness, the equations of motion obtained from the full Hamiltonian~\eqref{eq:H_nonlin} and the linearization around the semiclassical steady state are detailed here.
We start by adding dissipation to the Hamiltonian~\eqref{eq:H_nonlin} and obtaining the Langevin equations
\numparts
\begin{eqnarray}
    \dot{{c}} = -(\kappa+\rmi\Delta_{c}){c} -\rmi(\lambda+\mu_0{q}){a}-\rmi g_0{c}{q}+\eta_\phi+\sqrt{2\kappa}{c}_\inpt,\\
    \dot{{a}} = -(\gamma+\rmi\Delta_{a}){a} -\rmi(\lambda+\mu_0{q}){c}+\sqrt{2\gamma}{a}_\inpt,\\
    \dot{{p}} = -\gamma_\mrm{m} {p}-\omega_\mrm{m} {q}-\mu_0({a}^{\dagger}{c}+{c}^{\dagger}{a})-
        g_0{a}^{\dagger}{a}+{\xi},\\
    \dot{{q}} = \omega_\mrm{m}{p},
\end{eqnarray}
\endnumparts
where the cavity and dopant dynamics is expressed in the rotating frame with respect to the driving frequency;
moreover, we defined $\eta_\phi = \eta\rme^{-\rmi\phi}$.

Next, we separate each operator into its classical amplitude and quantum fluctuations, ${o} = \bar{o} + \delta o$.
The classical amplitudes obey the steady state equations
\numparts
\begin{eqnarray}
    -(\kappa+\rmi\Delta_{c})\bar{c}-\rmi(\lambda+\mu_0\bar{q})\bar{a}-\rmi g_0\bar{c}\bar{q}+\eta_\phi =0,\\
    -(\gamma+\rmi\Delta_{a})\bar{a}-\rmi(\lambda+\mu_0\bar{q})\bar{c} =0,\\
    -\omega_\mrm{m}\bar{q}-\mu_0(\bar{c}\bar{a}^*+\bar{a}\bar{c}^*)-g_0|\bar{c}|^2 =0.;
\end{eqnarray}
\endnumparts
the solutions are
\numparts
\begin{eqnarray}
    \bar{a} =-\rmi\frac{\lambda+\mu_0\bar{q}}{\gamma+\rmi\Delta_{a}}\bar{c},\\
    \bar{q} =-\frac{g_0|\bar{c}|^2-2\lambda\mu_0\Delta_{a}|\bar{c}|^2/(\gamma^2+\Delta_{a}^2)}
        {\omega_\mrm{m}-2\mu_0^2\Delta_{a}|\bar{c}|^2(\gamma^2+\Delta_{a}^2)},\\
    \eta_\phi =\left[\kappa+\rmi\Delta_{c}+\rmi g_0\bar{q}+\frac{(\lambda+\mu_0\bar{q})^2}{\gamma+\rmi\Delta_{a}}\right]\bar{c}.\label{eq:amulti}
\end{eqnarray}
\endnumparts
Introducing $g = g_0\bar{c}$ and $\mu = \mu_0\bar{c}$, we recast (\ref{eq:amulti}) as
\begin{eqnarray}\label{eq:SteadyState}\eqalign{
    \eta_\phi &=
    \Bigg[\kappa+\rmi\Delta_{c} -\rmi\frac{g^2-2g\lambda\mu\Delta_{a}/(\gamma^2+\Delta_{a}^2)}
        {\omega_\mrm{m}-2\mu^2\Delta_{a}/(\gamma^2+\Delta_{a}^2)} \\
        &\qquad+\frac{\mu^2}{\gamma+\rmi\Delta_{a}}
        \left(\frac{\omega_\mrm{m}-g\mu/\lambda}{\omega_\mrm{m}-2\mu^2\Delta_{a}/(\gamma^2+\Delta_{a}^2)}\right)^2\Bigg]\bar{c},}
\end{eqnarray}
the solution of which is the intracavity field amplitude $\bar{c}$, implicitly contained in $g$ and $\mu$.

Without dopant ($\lambda = \mu = 0$) one retrieves the usual dispersive Kerr bistability equation for the intracavity field
\begin{equation}
    \eta_\phi = \left(\kappa+\rmi\Delta_{c}-\rmi\frac{g^2}{\omega_\mrm{m}}\right)\bar{c}
\end{equation}
and, in the absence of a dynamical instability, the motion-induced nonlinear phase-shift leads to optical bistability when the Kerr dephasing is of the order of $\kappa$.
With dopant, however, the bistability threshold can be lowered (or highered) owing to interference between various terms in \eqref{eq:SteadyState}.
We assume here that the system is stable and a single solution $\bar{c}$ exists.

Linearized fluctuations around the steady state obey the Langevin equations
\numparts
\begin{eqnarray}
    \delta\dot{c} = -(\kappa+\rmi\Delta_{c})\delta{}c-\rmi\lambda \delta{}a-\rmi\tilde{g}\delta{}q+\sqrt{2\kappa}c_\inpt,\\
    \delta{}\dot{a} = -(\gamma+\rmi\Delta_{a})\delta{}a-\rmi\lambda \delta{}c-\rmi\mu \delta{}q+\sqrt{2\gamma}a_\inpt,\\
    \delta{}\dot{p} = -\gamma_\mrm{m}\delta{}p-\omega_\mrm{m}\delta{} q-\mu(\delta{}a+\delta{}a^{\dagger}) -\tilde{g}^\ast \delta{}c-\tilde{g}\delta{}c^{\dagger}+\xi,\\
    \delta{}\dot{q} = \omega_\mrm{m} \delta{}p,
\end{eqnarray}
\endnumparts
where, to simplify the notation, we absorbed the term $g\bar{q}$ into $\Delta_{c}$, redefined $\lambda$ to include the term $\mu_0\bar{q}$, introduced $\tilde{g} = g - \rmi\lambda\mu/(\gamma+\rmi\Delta_{a})$, and set the driving phase $\phi$ such that $\bar{c}\in\mathbb{R}$.
We can associate the coherent dynamics in these equations with the linearized Hamiltonian given in \eqref{eq:H_lin};
for simplicity of notation, we drop the $\delta$ in the linearized Hamiltonian~\eqref{eq:H_lin} and the following calculations from the operators.

\section{Lyapunov equation and dynamical stability}\label{app:Lyapunov}

The drift and diffusion matrices $\vect{A}$, $\vect{N}$ in the Lyapunov equation~\eqref{eq:Lyapunov} can be obtained from the Hamiltonian and the jump operators~\cite{Cernotik2015}.
For the Hamiltonian in \eqref{eq:H_lin}, the assumed decay of the cavity field and the dopant, and the thermal noise acting on the mechanical resonator, one gets
\numparts
\begin{eqnarray}\fl
    \vect{A} = \left(
    \begin{array}{cccccc}
        -\kappa & \Delta_{c} & 0 & \lambda & -\sqrt{2}\eta\gamma & 0\\
        -\Delta_{c} & -\kappa & -\lambda & 0 & -\sqrt{2}(g-\eta\Delta_{a}) & 0\\
        0 & \lambda & -\gamma & \Delta_{a} & 0 & 0\\
        -\lambda & 0 & -\Delta_{a} & -\gamma & -\sqrt{2}\mu & 0\\
        0 & 0 & 0 & 0 & 0 & \omega_\mrm{m}\\
        -\sqrt{2}(g-\eta\Delta_{a}) & -\sqrt{2}\eta\gamma & -\sqrt{2}\mu & 0 & -\omega_\mrm{m} & -\gamma_\mrm{m}
    \end{array}
    \right),\\ \fl
    \vect{N} = \mathrm{diag}[2\kappa,2\kappa,2\gamma,2\gamma,0,2\gamma_\mrm{m}(2\nbar+1)];
\end{eqnarray}
\endnumparts
here, we defined $\eta = \lambda\mu/(\gamma^2+\Delta_{a}^2)$.
The system remains dynamically stable if the real parts of all the eigenvalues of the drift matrix $\vect{A}$ are nonpositive.

\section*{References}

\end{document}